# Could a Hexagonal Sunspot Have Been Observed During the Maunder Minimum?

V.M.S. Carrasco[1] (ORCID: 0000-0001-9358-1219), J.M. Vaquero[2,3] (ORCID: 0000-0002-8754-1509), M.C. Gallego[1,3] (ORCID: 0000-0002-8591-0382)

[1] Departamento de Física, Universidad de Extremadura, 06071 Badajoz, Spain [e-mail: vmscarrasco@unex.es]

[2] Departamento de Física, Universidad de Extremadura, 06800 Mérida, Spain

[3] Instituto Universitario de Investigación del Agua, Cambio Climático y Sostenibilidad (IACYS), Universidad de Extremadura, 06006 Badajoz, Spain

**Abstract:** The Maunder Minimum was the period between 1645 and 1715 whose main characteristic was abnormally low and prolonged solar activity. However, some authors have doubted this low level of solar activity during that period by questioning the accuracy and objectivity of the observers. This work presents a particular case of a sunspot observed during the Maunder Minimum with an unusual shape of its umbra and penumbra: a hexagon. This sunspot was observed by Cassini in November 1676, just at the core of the Maunder Minimum. This historical observation is compared with a twin case that occurred recently in May 2016. The conclusion reached is that Cassini's record is another example of the good quality observations made during the Maunder Minimum, showing the meticulousness of the astronomers of that epoch. This sunspot observation made by Cassini does not support the conclusions of Zolotova and Ponyavin (*Astrophys. J.* 800, 42, 2015) that professional astronomers in the 17th century only registered round sunspots. Finally, a discussion is given of the importance of this kind of unusual sunspot record for a better assessment of the true level of solar activity in the Maunder Minimum.

## 1. Introduction

The photosphere is characterized by a granulation pattern of cells with a lifetime of around 5–10 minutes which are about 1000 km in diameter. This pattern is the manifestation of convective movements in the solar interior (Nordlund, 2003). Sunspots are dark regions that appear in the photosphere as a consequence of the interaction between strong magnetic fields emerging below the photosphere and the movement of the solar plasma (Solanki, 2003). A typical sunspot is characterized by a dark core called the umbra surrounded, partially or completely, by a fainter region called the penumbra (Bray and Loughhead, 1964). Some sunspots contain bright structures with a lane shape inside the umbra called light bridges. Generally, the penumbra occupies the largest area of a sunspot. Its shape is irregular and depends on the evolutionary stage of the sunspots, but its outer edge is always sharp. Radial structures can be distinguished in it consisting of elongated dark and bright filaments. The bright filaments consist of bright grains that drift towards the umbra, becoming increasingly like umbral dots, and the dark structures in the penumbra consist of radial filaments that sometimes overlie the grains. However, there seems to exist a dark background in the penumbra (Foukal, 2004). Furthermore, the temporal evolution of the sunspot number on the solar disk has a cyclic character: approximately each eleven years there occurs a maximum. This cycle is known as the 11-year or Schwabe solar cycle (Usoskin, 2017).

The Maunder minimum (MM) is the only grand minimum of solar activity registered during the telescopic era (Eddy, 1976). Therefore, it is a key event in understanding the behavior of the long-term solar activity and its influence on the heliosphere and the climate of our planet. Since the publication of the benchmark article by Eddy (1976), other works have described important characteristics of this period. Ribes and Nesme-Ribes (1993) showed that there was a strong hemispheric asymmetry (most sunspots were observed in the southern hemisphere). A flat and very low level of solar activity during the MM was found by Hoyt and Schatten (1998) using historical sunspot group counts. However, recent studies have changed some of these ideas about MM. Clette et al. (2014) showed that several observations corresponding to the MM compiled by Hoyt and Schatten (1998) were recovered from solar meridian observations, and thus the solar activity for this period could be underestimated (Vaquero and Gallego, 2014). Moreover, a 9-year solar cycle was obtained by Vaquero et al. (2015) from subsets of the Hoyt and Schatten database. Vaquero et al. (2016) have published a revised collection of sunspot group numbers corresponding to the entire telescopic period until 2010, including new recovered records and a revision of the previously available observations for the MM.

Zolotova and Ponyavin (2015) suggested that the MM was not a grand minimum of solar activity, since they derived values of the sunspot number up to 100. An argument used by Zolotova and Ponyavin (2015) was the idea that contemporary astronomers were omitting in the historical records the sunspots with irregular or non-circular shapes, inter alia, because they were expecting planetary transits. They suggested that this may have been caused by the dominant worldview of that time that spots are shadows from a transit of unknown celestial bodies. Thus, objects on the solar disk with an irregular shape would not have been registered, and therefore the solar activity during the MM have been underestimated. In this controversy about the true level of the solar activity during the MM, Usoskin et al. (2015) and Gómez and Vaquero (2015) have pointed to some important errors contained in that study. Furthermore, the typical values of the level of solar activity obtained by Carrasco, Villalba-Álvarez and Vaquero (2015) and Carrasco and Vaquero (2016) from contemporary records are compatible with a grand minimum of solar activity.

In any case, Zolotova and Ponyavin (2015) demonstrate the importance of the interpretation of the historical records used to study the MM, as well as the social, political, and religious circumstances of those records. In this work, we analyse an exceptional sunspot record by Cassini (1730). We compare it with a modern observation of a sunspot with a similar morphology. Our aim is to shed light on some MM issues by demonstrating the high quality of the observations made by some astronomers of that period.

## 2. Observations

In the historical sources, one can find several examples of drawings and descriptions of non-circular sunspots (Vaquero and Vázquez, 2009; Usoskin *et al*., 2015). In an appendix of his book *Selenographia*, Hevelius describes sunspot observations he made during the period 1642–1645. Those records report sunspots with different morphologies including some with an irregular shape. Other examples of non-circular sunspots registered in the 17th century can be found in the drawings made by Galileo Galilei (Vaquero, 2004), Christoph Scheiner (Arlt *et al*., 2016), Nicholas Bion (Casas, Vaquero, and Vázquez, 2006), Marcgraf (Vaquero *et al.*, 2011), Strażyc (Dobrzycki, 1999), La Hire (Ribes and Nesme-Ribes, 1993), and other observers (Usoskin *et al*., 2015). We here want to highlight a sunspot observed by Cassini around 28-29 November 1676 (Cassini, 1730). It can be clearly seen that Cassini drew this sunspot with a hexagonal shape (Figure 1). This sunspot appeared on the solar limb on 18 November, and its aspect changed throughout its transit across the solar disk (Figure 2). On the first days, 18 and 19, the foreshortening of the sunspot due to its proximity to the solar limb is obvious. On the following days, it takes on a more circular appearance until 28 and especially 29 November when it can be seen to have acquired a hexagonal shape. After 29 November, the foreshortening due to its closeness to the solar limb is again clear. Thus, this historical sequence of images illustrates the meticulous work Cassini did when recording these details that he observed during this sunspot's transit across the solar disk.

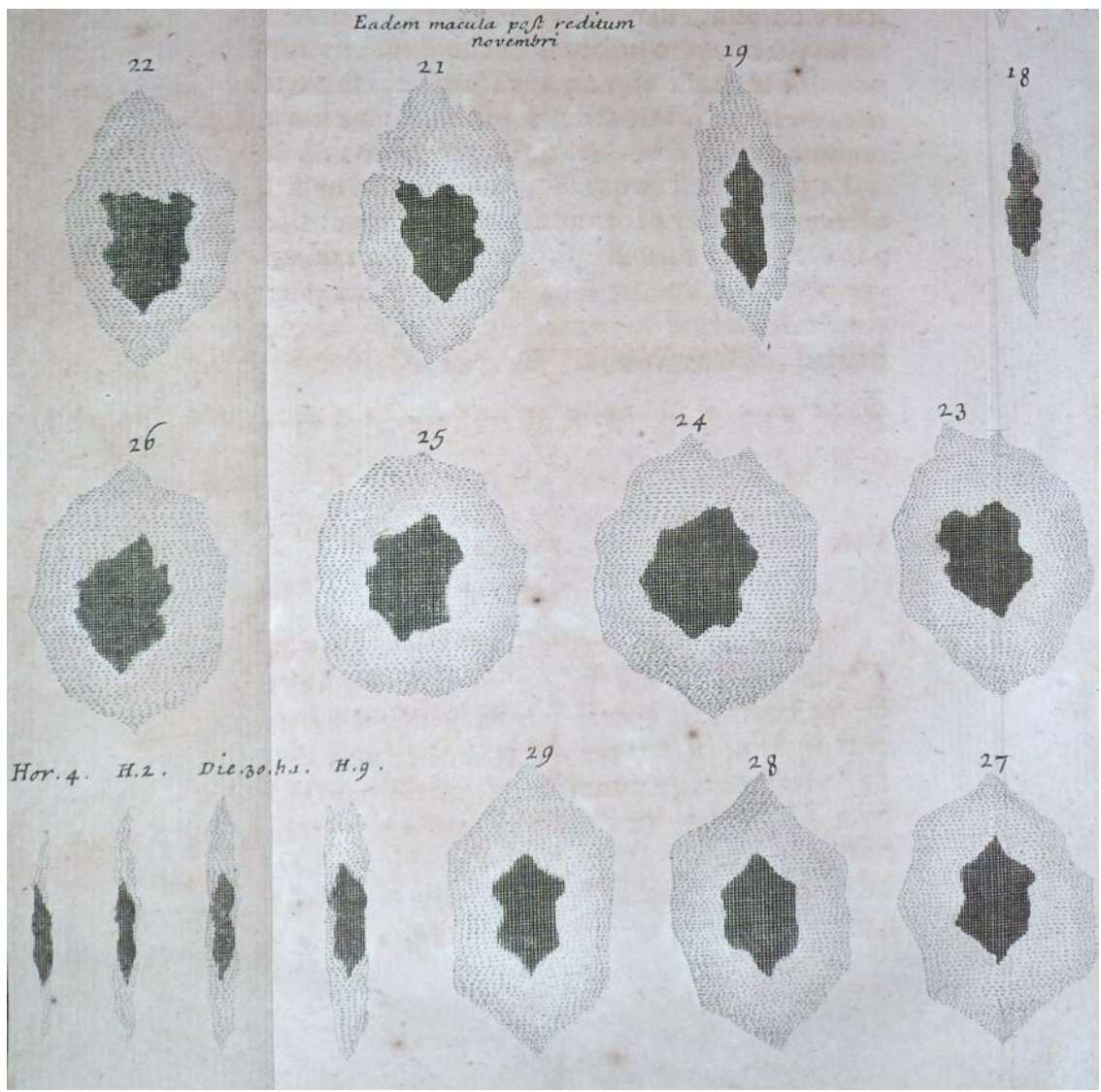

Figure 1. Sunspot observed by M. Cassini around 29 November 1676 (Cassini, 1730) [Courtesy of the Historical Archive and Library of the Real Instituto y Observatorio de la Armada, San Fernando (Cádiz), Spain].

A Hexagonal Sunspot During the Maunder Minimum?

This observed hexagonal shape is rather uncommon in the wide variety of sunspot morphologies. Therefore, the fidelity of this peculiar sunspot drawing to the actual sunspot shape may look questionable. However, we have found a modern example of a sunspot with a hexagonal appearance (Figure 3), observed around 22 May 2016, and designated AR12546 by NOAA. It was born on the visible hemisphere of the Sun on 26 April (at 7° S and 35° W) and was observed for the last time on 14 July (at 6° S and 23° W). It therefore had a lifetime of 80 days, and completed three synodic solar revolutions. In its second revolution, it appeared on the eastern solar limb on 14 May 2016 and disappeared from the western limb on 27 May 2016. On 21 May 2016, a C1.0 flare was produced by this active region, lasting 12 minutes and peaking at 14:00 UT. Its position on the solar disk at that moment was 9° S and 18° W. On 22 May, its position was 7° S and 35° W, its area was 550 millionths of a solar hemisphere, and was classified as Cho and β according to the McIntosh and Hale systems, respectively. In that transit across the solar disk, two more solar flares originated from that active region: i) one C1.3 type solar flare on 24 May 2016 at 10:16 UT, and ii) another C1.0 type solar flare on 26 May 2016 at 13:45 UT.

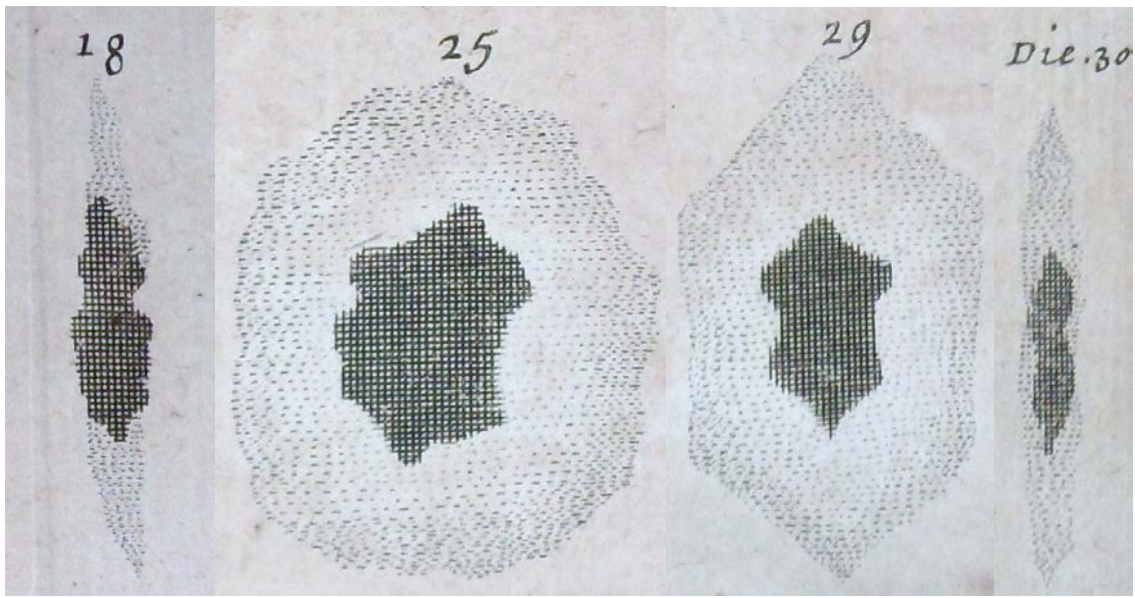

Figure 2. An enlargement of some parts of Figure 1 showing the evolution of the sunspot observed by Cassini in November 1676 (Cassini, 1730) [Courtesy of the Historical Archive and Library of the Real Instituto y Observatorio de la Armada, San Fernando (Cádiz), Spain].

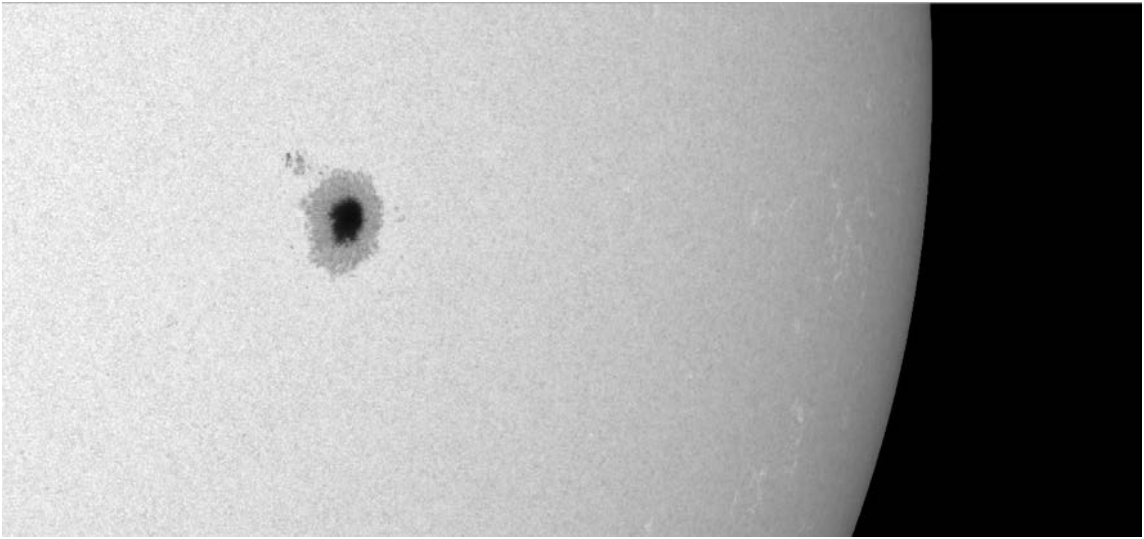

Figure 3. Sunspot with a hexagonal shape observed on 22 May 2016. This image was taken by the *Helioseismic and Magnetic Imager* (HMI) instrument (in continuum) on board the *Solar Dynamics Observatory* on 22 May 2016 at 23:28:22.700 [Source: https://sdo.gsfc.nasa.gov/].

## 3. Conclusions

Zolotova and Ponyavin (2015) concluded that solar activity during the MM is underestimated, arguing that any object on the solar disk with an irregular shape or consisting of a set of a small spots might have been omitted in textual reports. However, one can find several drawings and descriptions of sunspots observed during that period with morphologies different from a circular shape. For example, in *Selenographia*, Hevelius provides a detailed description of the morphology, size, or colour of sunspots observed between 1642 and 1645 which include different types of sunspots (Hevelius, 1647). We have here described two observations of a special case of sunspots with a hexagonal shape. The first was observed by Cassini in November 1676 during the MM, and the second in May 2016 with the most recent technology. After analysing these two cases, we must conclude that Cassini's unusual record of a hexagonal sunspot is an example of the good quality of the observations made by some astronomers during the MM. This and other examples shed doubts on the assumption of biased sunspot records and consequently, on the possibility that solar activity would be largely underestimated during the MM. Despite the social, political and religious circumstances of the 17th century scientists, which were very different from the current ones, they tried to capture in their drawings or reports the reality that they were observing with their telescopes. Therefore, their observations are trustworthy and scientifically usable. It is necessary to continue with the task of recovery and review of the sunspot records, especially those corresponding to early observations, with the objective of characterizing and understanding better the behavior of solar activity.

**Acknowledgements** All the historical materials used in this work were consulted at the Historical Archive and Library of the Real Instituto y Observatorio de la Armada, San Fernando (Cádiz), Spain. This research was supported by the Economy and Infrastructure Counselling of the Junta of Extremadura through project IB16127 and grant GR15137 (co-financed by the European Regional Development Fund) and by the Ministerio de Economía y Competitividad of the Spanish Government (AYA2014-57556-P and CGL2017-87917-P).

**Disclosure of Potential Conflicts of Interest** The authors declare that they have no conflicts of interest.